\documentclass[12pt]{article}

\usepackage {amsmath}
\usepackage {amssymb}
\usepackage {graphicx}

\begin{document}
\title {Cerebral cortex inspired representation of neural field network}

\author{Anil Kumar Sharma\\ Asha Sharma\\Birla Institute of Technology\\ Mesra, Ranchi - 814142, India}
\date{}  
\maketitle 

 \begin{abstract}
Evolution and its intelligence element present thrill and challenges in its exploration. Yet, how species have memory, retrieve them and maintain continuity are the fundamental questions.  Most of the phenomenon can only be hypothesised by researchers and validating them through experiments is a big challenge. Taking brain as an ideal intelligent machine and modelling it opens new dimensions for computational algorithm. This paper presents a hypothesis to resemble memory creation in cerebral cortex. The regions of cerebral cortex are implicit to be specific for specific function and constitute neural field that is single dimension and have vector form. The neural field throughout cortex connects with each other to form a network. These networks associate with survival instincts, emotions and rewards to constitute a memory of the exposed environment or say learning. Graphical tool NURBS with multidimensional control points are implicitly used in representing these networks as a set of cubic equations. Learning through data is a primary block of intelligent system, the paper attempts to convert the data in lower dimension patterns rather than existing absolute form for real time intelligent systems.
 \end{abstract}

\section{Introduction}
Basic instinct of ensuring safety, procuring energy and continuing reproduction driven by neural networks makes species to survive. Extended further, cognitive capabilities provide an edge to humans. Intensive research had finely elaborated neural network in human’s brain. Cerebral cortex is a major constitute of brain, it is a sheet of grey matter spanning equally in both right and left hemisphere with surface area of approximately $1000 cm^2$ in each. The cortex thickness ranges from 2 to 4 mm and have estimated 21 to 26 billion neurons. Brodmann in 1907 identified 52 distinct regions on cortex and further physiological investigations provides more with specific functions related to sensory, motor and association. Studies suggest that the neuron in a region is specific for a particular characteristic, like position cell, border cell and orientation cell in visual cortex.\\
The cerebral cortex covers and have network with thalamus, limbic system and other brain parts that drives survival instinct and emotions. Neural networks are formed by synaptic connections among neurons of various neural fields. Donald Hebb in 1949 given a famous quote for networking of neurons “Neurons that fire together, wire together”. In literature elaborations are there on connections at synapses for short as well as long term potentiation. Neurotransmitter as chemical messengers too have vital role in communication through network. \\
Brain performing complex task is an ideal outcome of evolution for researchers to develop intelligent systems. Simultaneously, remarkable development in mathematics and its applications, computational capacities, data storage, extraction and analysis, and algorithms opened new horizon for artificial intelligence (AI) and machine learning. The fundamental step for an artificial intelligent system is learning through data. Newer applications like natural language processing and generative AI are milestones in its development.\\
Brain’s intelligence and artificial intelligence processes have totally different approach. AI is an outcome of research and as such technologies are well understood whereas brain, an outcome of evolution, is mysterious. Available technologies can isolate and study every organ and region of brain, researchers had identify nearby function of each organ and region but yet understanding how collectively it associates with memory, sensory inputs, emotions and survival instinct is a challenge. Exploration of live Brain’s functions is a multidisciplinary aspect similar to that of space where we have extensive research scope.

\section{Problem statement}
Computational memory is an addressable absolute data stored physically in transistors as potential difference at binary level say 0 or 1 whereas brain’s memory is not similar. Primarily understanding and representing brain’s memories process is useful for information storage, retrieval, association and continuity in real time intelligent system.\\
The term “absolute” in problem statement means that the saved information is definite in nature on retrieving at any time. Like if someone asked for number of stairs in a familiar place may not be able to recall but if the picture of stairs is shown the exact count is known. This paper hypothesizes memory creation process and attempts to represent it by a general graphical tool.

\section{Hypothesis in framing the model}
Brain creates memories as network of neural field associated with survival instincts, emotions and rewards during sensory exposure to environment or say learning. The cerebral cortex stores such networks by sampling environment on stimulus. Here the stimulus means the derivative passing threshold, tends to take sample of environment at that instant. These stimuli are from senses, survival instinct, emotions and rewards. These networks of neural filds represents extracted features of exposed environment at lower dimensions and invariantly retrivable. 

\section{Mathematical representation}

\subsection{Neural field}

A localized column in cerebral cortex and other region specialized for specific function. Each and similar specialized neural field represents single dimension. These dimensions are defined as sets, subsets and the subset of their respective subsets. Primary sets are Sensory (SE), Survival Instinct (SI), Emotions (EM), Motor (MO) and Reward (RE) their subsets are as follows.\\
%\begin{table}
A. Sensory subsets\\
1.	SEA - Sensory Auditory
2.	SEV - Sensory Visual
3.	SES - Somatosensory
4.	SEG – Gustatory Sensory
5.	SEO -Olfactory sensory
6.	SEB - Vestibular Sensory\\
B. Survival Instinct subsets\\
1.	SIS – Safety
2.	SIE – Energy
3.	SIR - Reproduction\\
C. Emotion subsets\\
1.	EMF - Fear
2.	EMA - Anger
3.	EMH - Happiness
4.	EMS – Sadness\\
D. Motor subsets\\
1.	MOC – Conscious voluntary movement
2.	MOS - Subconscious Muscle memory
3.	MOI - Involuntary reflexes\\
E. Reward subsets\\
1.	RED - Dopamine
%\end{table}
\\
The set of dimensions expressed in vector form as:\\

$
\begin {bmatrix}
SEA&-&SIS&-&EMF&-&MOC& - &RED \\
\end {bmatrix}
$

\subsection{Network}
A network is a group of neural field in lower dimensional manifold of higher dimension exposed environment. All associated neural fields of a network fires together. Figure 1 outline flow diagram for network formation. As long as the brain is exposed to environment and receives stimulus it keep adding and reviving network.\\

\begin{figure}
  \includegraphics[width=\linewidth]{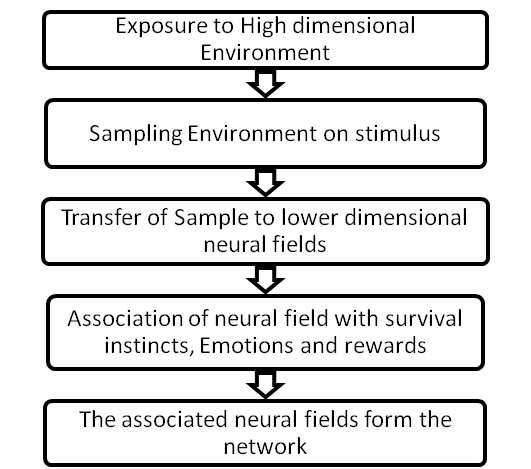}
  \caption{Flow diagram for network formation}
  \label{fig:Flow diagram for network formation}
\end{figure}

In general network is connecting control points. For visualization, we take points on two dimensional surfaces. These points can be connected by linear spline, Cardinal spline or Catmull-Rom spline that interpolate each point but have $C^0$ continuity. Continuity aspects are considered for continumn progression as transition from one network to another. Bezier and Hermite curves with $C^1$ continuity. The most commonly used curves are Basis Spline (B-spline) having $C^2$ continuity. Varying the knot interval of B-spline, they became non uniform and including weightage to the control points make them rational, commonly known as Non Uniform Rational B-Spline curve (NURBS).\\

$K_{th}$ Degree NURBS curve is defind by\\

$C(u) = \frac{\sum^n_{i=0} N_{i,k}w_iP_i}{\sum^n_{i=0} N_{i,k}(u)w_i}$\\

The input u value is one dimensional space called parameter space.\\
The points $P_i$ are defined as control points.\\

$N_{i,k}(u)$ are $k_{th}$ degree B-Spline basis function defined on the non uniform knot vector.\\

$w_i$ are weights, here, $w_(i )> 0$ for all i\\

$ u = ( a,.....a,u_{k+1},........u_{n-k-1},b,.....,b)$\\ 

$a\leq u \leq b$\\

For rational piecewise\\

$R_{i,k}(u) = \frac{N_{i,k}(u) w_i}{\sum^n_{j=0} N_{j,k}(u)w_j}$\\

For all  u $  \epsilon [0,1]$\\

With properties of positivity: $R_{i,k}(u) \geq 0$ for all i, k and u\\
 and, partitin of unity: $\sum^n_{i=0} R_{i,k} (u) = 1$\\

Control point in a network is multi-dimensional and every dimension is a vector quantity. Three degree cubic B-spline curve need four control points, that make the control point matrix (P) size of 4 X number of dimensions.
\\

P =
$
\begin {bmatrix}
SEA_0&-&SIS_0&-&EMF_0&-&MOC_0& - &RED_0 \\
SEA_1&-&SIS_1&-&EMF_1&-&MOC_1& - &RED_1 \\
SEA_2&-&SIS_2&-&EMF_2&-&MOC_2& - &RED_2 \\
SEA_3&-&SIS_3&-&EMF_3&-&MOC_3& - &RED_3 \\

\end {bmatrix}
$
\\
\\
Characteristic matrix (M) for B-spline basis function is;
\\
M=
$
\begin {bmatrix}

1&4&1&0\\
-3&0&3&0\\
3&-6&3&0\\
-1&3&-3&1\\

\end {bmatrix}
$
\\
\\Parametric space (U) in cubic equation matrix is; 
\\
U=
$
\begin {bmatrix}

1&u&u^2&u^3\\

\end {bmatrix}
$
\\

The cubic equation for C(u) is solved as matrix product:\\

C(u) = [U] [M][P]
\\
The solution of above results in cubic equations equal to the number of dimension. These cubic equation represents the neural field network for the relevant enviroment exposure and constitutes memory as patterns.

\section{Discussion}

Hypothesizing neural network and representing them through multidimensional NURBS is an attempt to view data in lower dimension. As a building block, features from exposed environment are required to identify in a vector form. Such features of sensory input and survival instinct can be generalized but emotions and rewards are dynamic and vary in perception. The network creation as discussed in article is next in the sequence. The control points of exposure are identified on vector scale of each dimension. These control points are transformed into a pattern represented by cubic equation. The set of cubic equation for an environmental exposure or learning is the set of neural field network.  On stimulus of similar control point the set of neural field network respond as outcome.\\
The validation and real time application of discussed hypothesis and representation require intense backward as well as forward integration of techniques though this is a shift in paradigm to view AI in future, inspired by naturally evolved intelligence.

\section{References}

\begin {enumerate}
\item Bhaduri, A., Sandoval-Espinosa, C., Otero-Garcia, M. et al. An atlas of cortical arealization identifies dynamic molecular signatures. Nature 598, 200–204 (2021). https://doi.org/10.1038/s41586-021-03910-8.
\item Cadwell CR, Bhaduri A, Mostajo-Radji MA, Keefe MG, Nowakowski TJ. Development and Arealization of the Cerebral Cortex. Neuron. 2019 Sep; 103(6):980-1004. DOI: 10.1016/j.neuron.2019.07.009. PMID: 31557462; PMCID: PMC9245854.
\item David J. Heegera. Theory of cortical function. PNAS. 2017; 114(8):1773–1782.
\item Edvard I Moser, Emilio Kropff, May-Britt Moser. Place Cells, Grid Cells, and the Brain’s Spatial Representation System. Annu. Rev. Neuroscience, 2008; 31: 69-89.
\item John O'Keefe, Lynn Nadel. The Hippocampus as a Cognitive Map, Oxford University Press. 1978.
\item Mark D Lescroart, Jack L Gallant. Human Scene-Selective Areas Represent 3D Configurations of Surfaces. Neuron.2018; 101:1-15. 
\item Martnez Canada P, Ness TV, Einevoll GT, Fellin T, Panzeri S. Computation of the electroencephalogram (EEG) from network models of point neurons. PLOS Computational Biology.2021;17(4):1-41.
\item Nunez Elizalde, A O. Using multiple high-dimensional feature spaces to model brain activity recorded during naturalistic experiments. UC Berkeley. 2018 Merritt ID: ark:/13030/m5vj1d00.
\item Rosario Tomasello, Max Garagnani, Thomas Wennekers, Friedemann Pulvermüller. A Neurobiologically Constrained Cortex Model of Semantic Grounding with Spiking Neurons and Brain-Like Connectivity. Frontiers in Computational Neuroscience. 2018;12:88.
\item Šimić G, Tkalčić M, Vukić V, Mulc D, Španić E, Šagud M, Olucha-Bordonau FE, Vukšić M, R Hof P. Understanding Emotions: Origins and Roles of the Amygdala. Biomolecules. 2021 May 31; 11(6):823. doi: 10.3390/biom11060823. PMID: 34072960; PMCID: PMC8228195.
\item Suzana Herculano Houzel. The human brain in numbers: a linearly scaled-up primate brain. Fronteiers in human neuroscience.2009; 3:31:1-11.
\item T. Triffet, H. S. Green.  Mathematical modeling of the cortex. Mathematical Modeling. 1984; 5:383-399.
\item Wandell BA, Dumoulin SO, Brewer AA. Visual field maps in human cortex. Neuron. 2007; 25:56(2):366-383.
\item Zlotnik G, Vansintjan A. Memory: An Extended Definition. Front Psychol. 2019 Nov 7; 10:2523. doi: 10.3389/fpsyg.2019.02523. PMID: 31787916; PMCID: PMC6853990.

\end {enumerate}

\end {document}